\documentclass[reprint,amsmath,nofootinbib,amssymb,aps,prd]{revtex4-2}
\usepackage{graphicx}
\usepackage{dcolumn}
\usepackage{bm}
\usepackage{hyperref}
\usepackage{orcidlink}
\newcommand{\be}{\begin{equation}}
\newcommand{\ee}{\end{equation}}
\usepackage{xcolor}
\definecolor{DarkGreen}{rgb}{0,0.6,0}

\begin{document}

\title{Could We Be Fooled about Phantom Crossing?}
\author{Ryan E.\ Keeley\orcidlink{0000-0002-0862-8789}}
  \email{rkeeley@ucmerced.edu}
 \affiliation{University of California, Merced, 5200 N Lake Road, Merced, CA 95341, USA}
 \author{Arman Shafieloo\orcidlink{0000-0001-6815-0337}}
 \email{shafieloo@kasi.re.kr}
 \affiliation{Korea Astronomy and Space Science Institute, 776, Daedeokdae-ro, Yuseong-gu, Daejeon 34055, Republic of Korea}
 \affiliation{University of Science and Technology, 217 Gajeong-ro, Yuseong-gu, Daejeon 34113, Republic of Korea}
  \author{William L. Matthewson\orcidlink{0000-0001-6957-772X}}
  \email{willmatt4th@kasi.re.kr}
 \affiliation{Korea Astronomy and Space Science Institute, 776, Daedeokdae-ro, Yuseong-gu, Daejeon 34055, Republic of Korea}

\date{\today}%

\begin{abstract}
Recent data from DESI Year 2 BAO, Planck CMB, and various supernova compilations suggest a preference for evolving dark energy, with hints that the equation of state may cross the phantom divide line ($w = -1$). While this behavior is seen in both parametric and non-parametric reconstructions, comparing reconstructions that support such behavior (such as the best fit of CPL) with those that maintain $w>-1$ (like the best fit algebraic quintessence) is not straightforward, as they differ in flexibility and structure, and are not necessarily nested within one another. Thus, the question remains as to whether the crossing behavior that we observe, suggested by the data, truly represents a dark energy model that crosses the phantom divide line, or if it could instead be a result of data fluctuations and the way the data are distributed. We investigate the likelihood of this possibility. For this analysis we perform 1,000 Monte Carlo simulations based on a fiducial algebraic quintessence model. We find that in $3.2 \% $ of cases, CPL with phantom crossing not only fits better, but exceeds the real-data $\chi^2$ improvement. This Monte Carlo approach quantifies to what extent statistical fluctuations and the specific distribution of the data could fool us into thinking the phantom divide line is crossed, when it is not. Although evolving dark energy remains a robust signal, and crossing $w=-1$ a viable phenomenological solution that seems to be preferred by the data, its precise behavior requires deeper investigation with more precise data.
\end{abstract}

\maketitle

\section{Introduction}
The accelerated expansion of the Universe remains one of the most profound mysteries in modern cosmology. Dark energy (DE), a hypothesized component with negative pressure, is the leading explanation for this phenomenon. Although the simplest model, a cosmological constant ($\Lambda$) characterized by a constant equation-of-state parameter $w = -1$, is well supported, it also suffers from well-known theoretical challenges, such as the fine-tuning and coincidence problems. This has motivated extensive exploration of dynamical dark energy models, in which $w$ varies with redshift. A widely used and flexible phenomenological framework is the Chevallier-Polarski-Linder (CPL) parametrization, defined as $w(z) = w_0 + w_a (1 - a)$, where $a$ is the scale factor ~\cite{CHEVALLIER_2001,Linder_2003}. This form allows for both quintessent ($w > -1$) and phantom ($w < -1$) behavior, including the possibility of crossing the so-called phantom divide line at $w = -1$.

Recent results from the Dark Energy Spectroscopic Instrument (DESI) have significantly improved constraints on the expansion history of the Universe. When DESI baryon acoustic oscillation (BAO) measurements are combined with cosmic microwave background (CMB) data and observations of Type Ia supernovae (SN), the resulting dataset exhibits a clear preference for a dynamical dark energy component over $\Lambda$CDM. Specifically, DESI analyses find that the CPL model, with best-fit parameters around $w_0 \approx -0.7$ and $w_a \approx -1.0$, is favored over $\Lambda$CDM at greater than $3-4\sigma$ significance, depending on the SN dataset used ~\cite{DESI:2024mwx,DESI:2025zgx,DESI:2024kob,DESI:2025fii}. Notably, these parameters imply that the dark energy component transitions from a quintessent phase at low redshifts to a phantom phase at higher redshifts, an evolution referred to as \emph{phantom crossing}.

This phantom crossing behavior is not suggested only by parametric models like CPL, but also emerges in several non-parametric reconstructions of the dark energy equation of state. Techniques such as Gaussian processes, Crossing statistics, binning methods and shape function reconstruction have independently indicated similar trends in the evolution of $w(z)$, lending additional weight to the possibility of dynamical dark energy behavior~\cite{DESI:2025fii,DESI:2025wyn}. However, these non-parametric approaches are also highly flexible by design and can more easily accommodate statistical fluctuations or systematic features in the data. As such, while they corroborate the CPL-inferred crossing, their agreement may be more reflective of their capacity to fit noise or small-scale variations rather than a robust underlying signal.

While the statistical preference for phantom-like evolution is intriguing, the physical implications are nontrivial. Phantom behavior, particularly in single-field scalar theories, generally entails the violation of the null energy condition (NEC), which can lead to instabilities such as ghost modes or superluminal propagation~\cite{Caldwell:1999ew,Valiviita:2008iv,Clifton:2011jh,Tsujikawa:2013fta,Hawking_Ellis_2023}. Consequently, some argue that the preference for phantom behavior might simply arise from the flexibility of certain parametrizations, instead of being a true indication of the fundamental physics of dark energy.

In parallel, growing theoretical and observational interest has turned toward non-phantom dynamical models, particularly those rooted in quintessence. These include thawing models, in which the scalar field slowly evolves away from a cosmological constant–like state, and models with specific potential forms (e.g., exponential or power-law potentials) that remain strictly in the $w > -1$ regime. Several recent works show that such models can provide fits to DESI, CMB, and SN data that are preferred over $\Lambda$CDM and competitive with CPL-based models that permit phantom crossing ~\cite{Shlivko:2024llw,Payeur:2024dnq,shlivko2025,Wolf:2025jlc,Akrami:2025zlb}. Furthermore, these models possess theoretical advantages, including stability and consistency with known field-theoretic principles.

Given this context, a pressing question arises: Is the preference for phantom crossing in observational data robust, or is it sensitive to the choice of parametrization and priors? More fundamentally, does the data actually \emph{require} crossing $w = -1$, or can it be adequately explained by physically motivated quintessence models that avoid such transitions? Since both parametric and non-parametric reconstructions can introduce significant modeling freedom, it is essential to discriminate between real physical signals and artifacts from overly flexible fitting.

In this work, we critically examine the observational evidence for phantom crossing by directly comparing the CPL model to a parametric form meant to represent non-phantom, physically motivated quintessence models. We use the latest DESI BAO data, in combination with compressed Planck CMB data and the Union3 SN compilation, to quantify the statistical significance of phantom crossing and assess whether simpler, theoretically well-behaved models can still provide a viable description despite their poorer fit to the data. Our analysis also explores the role of model complexity and interpretability of fitted parameters, with the goal of understanding whether current data genuinely point to exotic dark energy behavior or whether standard, non-pathological alternatives to $\Lambda$ remain viable.

\section{Methodology}

\subsection{Models/Parametrizations}
To explore the potential phantom crossing behavior of dark energy, we consider two distinct parametrizations of the dark energy equation-of-state parameter $w(z)$, each with strengths in capturing particular redshift evolutions. These two models are not nested, meaning neither is a subset of the other, but both contain the standard $\Lambda$CDM limit within their parameter spaces. The focus is not simply on the parametric form itself, but on the ability of each model to reflect actual features suggested by the data.

The first model we consider is the widely-used CPL parametrization, often referred to as the $w_0w_a$ model \cite{CHEVALLIER_2001, Linder_2003, de_Putter_2008}. This parametrization expresses $w(z)$ as:
\begin{equation}
    w(z) = w_0 + w_a\frac{z}{1+z},
\end{equation}
where $w_0$ represents the present-day value of the equation of state and $w_a$ determines its evolution with redshift. The CPL form has the advantage of simplicity and flexibility, able to emulate a wide variety of underlying physical models. Notably, it has been shown to effectively capture observable features arising from a broad class of scalar field models \cite{de_Putter_2008}. CPL is the primary parametrization used in the DESI DR2 analysis, which favors a phantom crossing behavior ($w < -1$ at some redshift), lending motivation for its continued study.

\begin{figure}
    \centering
    \includegraphics[width=\linewidth]{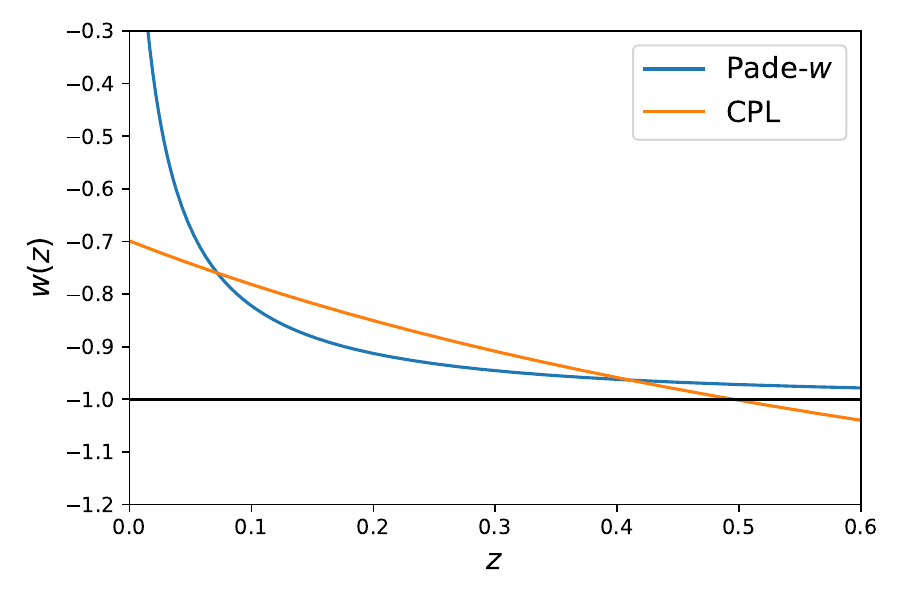}
    \caption{Best-fit $w(z)$ functions for both the CPL (orange) and Pade-$w$ (blue) models. The phantom divide line ($w = -1$) is shown in black.}
    \label{fig:exCPL_Pade}
\end{figure}

The second model is the Pad\'e approximant parametrization, often called Pade-$w$, which was developed to better reflect the general dynamics of thawing quintessence models \cite{alho2024, shlivko2025}. The Pad\'e parametrization for $w(z)$ takes the form:
\begin{equation}
    w(z) = \frac{2\epsilon_0}{3 + \eta_0(z^3 + 3z^2 + 3z)} - 1,
\end{equation}
written in terms of two physically interpretable parameters\footnote{For the physical interpretations see for e.g. \cite{shlivko2025}}: $\epsilon_0$, which sets the present-day value of $w(z)$, and $\eta_0$, which governs the rate of redshift evolution; the high-redshift limit of $w(z)$ in the Pade-$w$ model is $w=-1$, so $\eta_0$ sets how fast $w(z)$ reaches $-1$. Compared to CPL, Pade-$w$ is particularly well-suited for modeling behaviors that naturally arise in field-based quintessence theories and offers a different perspective on the underlying physics.  In this analysis we also restrict ourselves to positive values of $\epsilon_0$ and $\eta_0$ which maintain non-phantom behaviour, $w(z)>-1$. As an example, we show the best-fit $w(z)$ behaviors for the two models in Fig.~\ref{fig:exCPL_Pade}.

While both parametrizations capture the $\Lambda$CDM behavior in specific limits, they respond differently to trends in the data. For example, the CPL model can easily emulate phantom-crossing behaviors, while the Pade-$w$ form better captures smooth thawing evolutions. This makes their comparison important, especially when trying to interpret whether the preference for $w < -1$ is a robust feature of the data or reconstruction-induced.

\subsection{Datasets}

Our analysis employs several state-of-the-art cosmological datasets that probe the expansion history of the Universe over a wide range of redshifts. These datasets are also used to generate mock realizations for statistical inference, as discussed in Sec.~\ref{Mocks}.\\ 

\textbf{CMB:} We use a compressed representation of cosmic microwave background (CMB) constraints, based on summary statistics used in \cite{DESI:2025zgx}. Specifically, we adopt Gaussian priors on $\theta_*$ (the CMB acoustic scale), $\omega_{bc}$ (physical baryon plus cold dark matter density, without neutrino contribution), and $\omega_b$ (physical baryon density). This type of approach enables efficient generation of mock datasets and has been shown to capture the essential CMB constraints on the late-time expansion history, with minimal loss of information compared to full-likelihood approaches \cite{Planck:2018vyg,Wang_2007,Lemos_2023}.

\textbf{SNe Ia:} For supernova data, we use the Union3 compilation \cite{Rubin:2023ovl}, which is based on 2087 spectroscopically confirmed Type Ia supernovae. The analysis uses 22 distance modulus nodes derived through a spline interpolation scheme, implemented using the Unity 1.5 Bayesian hierarchical framework. These measurements cover a redshift range of $0.05 < z < 2.26$, offering robust constraints on the shape of the distance-redshift relation.

\textbf{BAO:} The DESI DR2 dataset \cite{DESI:2025zgx} provides precise BAO measurements. This includes determinations of the Hubble distance $D_H(z)$ and angular diameter distance $D_M(z)$ across six redshift bins spanning $0.51 < z < 2.33$, as well as an additional measurement of the isotropic distance scale $D_V(z)$ at $z = 0.295$. The DESI data significantly improve constraints on the intermediate- to high-redshift behavior of $w(z)$.

\subsection{Statistics}
Since the CPL and Pade-$w$ models are not nested, a direct Bayesian comparison is not straightforward. In nested models, one can calculate posterior distributions for a parameter that transitions between models, and assess preference based on whether the value of the parameter for the simpler model lies within a high-probability region. In our case, no such bridging parameter exists. Moreover, using the Bayesian evidence ratio to compare models is sensitive to the choice of prior ranges, which can significantly influence the final outcome.

Thus, we adopt a frequentist approach to assess model preference. Specifically, we use the best-fit Pade-$w$ model as a fiducial model to generate mock data realizations, which allows us to evaluate the performance of each model in a controlled environment. For each mock realization, we fit both the CPL and Pade-$w$ models and compare their $\chi^2$ values. By constructing the distribution of $\Delta\chi^2$ across the ensemble of realizations, we determine the statistical likelihood of spuriously obtaining a preference for CPL, simply due to statistical fluctuations.

This enables us to compute the Type I error rate (or $\alpha$-value), the probability of falsely favoring CPL when Pade-$w$ is the true underlying model. This method circumvents the difficulties of Bayesian evidence computation for non-nested models and is especially well-suited for testing significance in high-dimensional parameter spaces.

\subsection{Generating Mock Data}\label{Mocks}

To quantify the significance of the apparent CPL preference in current data, we generate mock datasets under the assumption that the Pade-$w$ model represents the true cosmology. This allows us to assess how often CPL could outperform Pade-$w$ due to statistical fluctuations alone.

First, we obtain the best-fit cosmological parameters for the Pade-$w$ model using the real DESI DR2, Union3, and compressed CMB datasets. Using these parameters, we compute the necessary theoretical predictions for key observables: the Hubble parameter $H(z)$, angular diameter distance $D_A(z)$, and distance modulus $\mu(z)$, evaluated at the redshifts of the actual data points. Together, with the best-fit values of the CMB compressed parameters, these form a fiducial dataset for the true model.

We then perturb these predictions with Gaussian noise drawn from the original data covariance matrices, thereby producing realistic mock datasets. This procedure is repeated multiple times (typically $\mathcal{O}(10^3)$) to generate an ensemble of mock realizations. For each mock dataset, we perform a full parameter estimation for both the CPL and Pade-$w$ models, allowing for a direct comparison of model fits.

The resulting distribution of $\chi^2$ differences over each realization in the ensemble yields a robust estimate of the expected statistical variation. By comparing the observed $\Delta\chi^2$ from real data to this distribution, we assess how atypical the observed CPL preference is under the null hypothesis that the true model is Pade-$w$.

This methodology offers a clean, interpretable framework for assessing the robustness of phantom-crossing claims, grounded in realistic assumptions about data quality and model flexibility.

\section{Results}

The first result we obtain is that, for the real data, CPL is preferred over the Pade-$w$ model at a $\Delta \chi^2 = 3.3$. We find best-fit parameter values of $w_0, w_a$ = $-0.7,-1.0$ for CPL and $\eta_0, \epsilon_0$ = $59,1.9$ for Pade-$w$. To remind the reader, we use this Pade-$w$ best-fit model to generate the mock data that we use to assess significance.

\begin{figure}
    \centering
    \includegraphics[width=\linewidth]{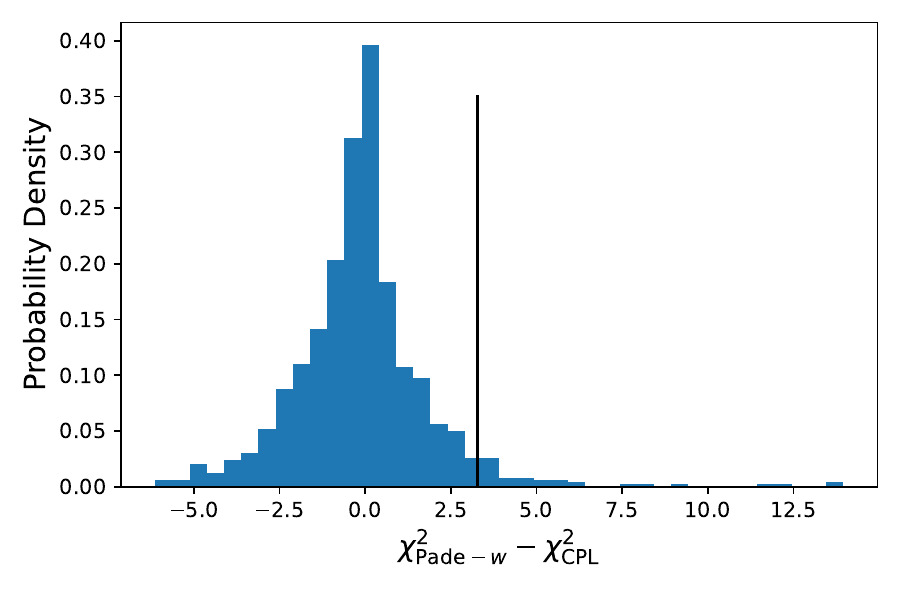}
    \caption{Distribution (blue) of the difference in $\chi^2$ values between the best-fit Pade-$w$ model and the best-fit CPL model fit to different realizations of mock data generated from the best-fit Pade-$w$ model to the real DESI DR2 + Union3 + CMB data. The black line is the difference in the best-fit $\chi^2$ value between the best-fit Pade-$w$ model and the best-fit CPL model to the real DESI DR2 + Union3 + CMB data. The preference for CPL over Pade-$w$ in the actual data occurs at that level or higher in 3\% of mock cases.}
    \label{fig:dchi2_dist}
\end{figure}

Next, we separately fit both the CPL model and the Pade-$w$ model to the different mock realizations of the CMB+DESI+Union3 data and plot the distribution of $\Delta\chi^2 \equiv \chi^2_{{\rm Pade}-w} - \chi^2_{CPL}$ values in Figure~\ref{fig:dchi2_dist}.
The vertical black line shows the difference in the $\chi^2$ value between the best-fit Pade-$w$ model and the best-fit CPL model to the real DESI DR2 + Union3 + CMB data. This level of preference, or higher, for CPL over Pade-$w$ in the actual data occurs in 3\% of mock cases, even though all the mock data are generated using the best-fit Pade-$w$ as the true model. This is our key result.

\subsection{Investigating why CPL is sometimes better}

The tautological explanation for this chance of preferring CPL over Pade-$w$, given mock data based on a Pade-$w$ fiducial model, is that CPL is just more flexible than Pade-$w$ and so has more of a chance of over-fitting than Pade-$w$. In the rest of this paper, we explore what features of the CPL model, and of the mock data, account for this spurious preference for CPL.  This will be useful for identifying which data will be most useful for testing whether quintessent or phantom crossing models are preferred, and where the effects of the limitations of current data precision are most important.

\begin{figure}
    \centering
    \includegraphics[width=1.1\linewidth]{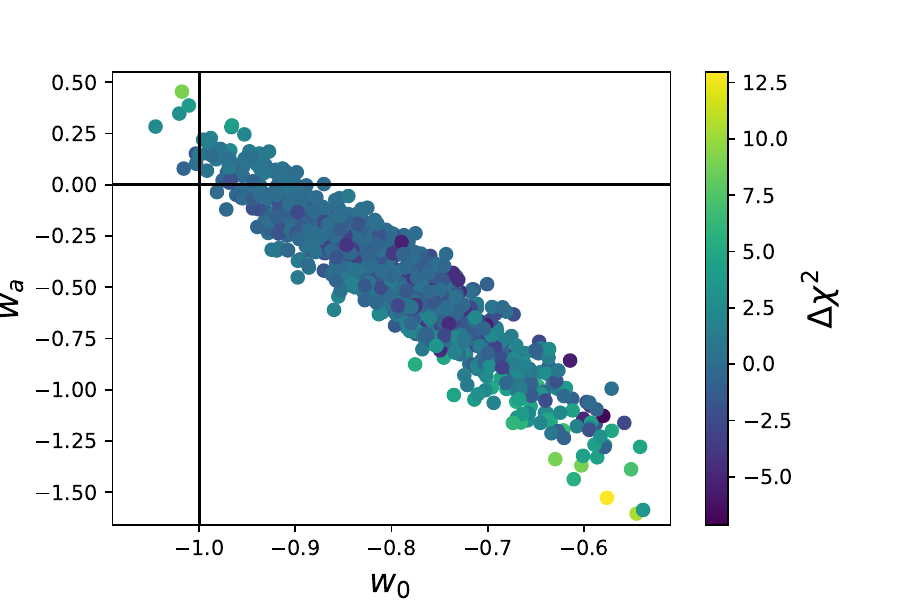}
    \caption{Distribution of best-fit CPL parameters fit to mock Pade-$w$ datasets, color-coded by the difference in $\chi^2$. }
    \label{fig:w0wa_dist}
\end{figure}

In Figure~\ref{fig:w0wa_dist}, we show the full distribution of best-fit CPL parameters obtained from fits to mock Pade-$w$ datasets, color-coded by the difference in $\chi^2$ with respect to the corresponding best-fit Pade-$w$ of the same mocks. Clearly, for the chosen datasets and true model, there is a preference for values of $w_0, w_a$ in the lower right quadrant. 
The largest values of $\Delta\chi^2$ appear to be situated on the extreme ends of the resulting region. 363 of these samples have both a better fit to the mock data than Pade-$w$ and a phantom crossing.

\begin{figure}
    \centering
    \includegraphics[width=1.1\linewidth]{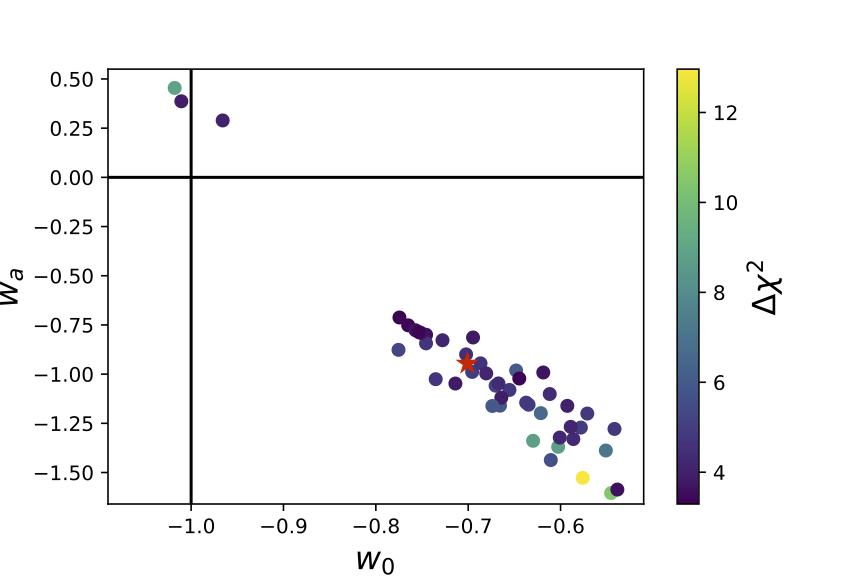}
    \caption{Distribution of best-fit CPL parameters fit to mock Pade-$w$ datasets, where the CPL model is a better fit than the Pade-$w$ model by a degree greater than that in the real data. The points are color-coded by the difference in $\chi^2$. }
    \label{fig:w0wa_dist_type2}
\end{figure}

In Figure~\ref{fig:w0wa_dist_type2} we show the distribution of best-fit CPL parameters fit to the mock Pade-$w$ datasets, only in the cases where the CPL model is a better fit than the Pade-$w$ model by a degree greater than that in the real data. The points are color-coded by the difference in $\chi^2$ and display a bimodal distribution, where most of the points are in the same region of parameter space that is preferred by the real DESI data and another region where $w_a$ is positive.

We now consider in more detail the most extreme (largest $\Delta\chi^2$) case of these realizations that spuriously prefer CPL over Pade-$w$. This will be useful to highlight at what redshifts CPL and Pade-$w$ make the biggest differences in their predicted $H(z)$ values, and thus at what redshifts it is most important to measure precisely in order to distinguish the two models.

\begin{table}
    \centering
    \begin{tabular}{lllll}
        \hline
    \hline
        $\chi^2$ & CMB & BAO & SN & Total \\
            \hline
    \hline
        Fiducial & 2.82 & 25.38 & 22.73 & 50.93\\
        \hline
        Pade-$w$   & 1.85 & 22.48 & 22.30 & 46.63\\
        \hline
        CPL      & 0.52 & 11.14 & 21.02 & 32.68\\
    \end{tabular}
    \caption{$\chi^2$ values for the mock dataset where CPL is most preferred over Pade-$w$. Each column is the $\chi^2$ value for the mock CMB, BAO, and SN dataset corresponding to each of the fiducial Pade-$w$, best-fit Pade-$w$, and best-fit CPL models in each row. }
    \label{tab:mock_chi2}
\end{table}
Table~\ref{tab:mock_chi2} shows that, for this mock, the BAO dataset is the most sensitive to differences in the observables of the Pade-$w$ model and the CPL model and so further improvements to the DESI likelihood (more precise measurements, and more measurements at higher redshift) will be necessary to be able to conclusively discriminate between purely quintessent models and models that have a phantom crossing, like CPL. This makes sense since BAO directly measure $H(z)$ which is one derivative closer to $w(z)$ than measurements of $D_L(z)$, which is all that SN can measure.

\begin{figure}
    \centering
    \includegraphics[width=1.1\linewidth]{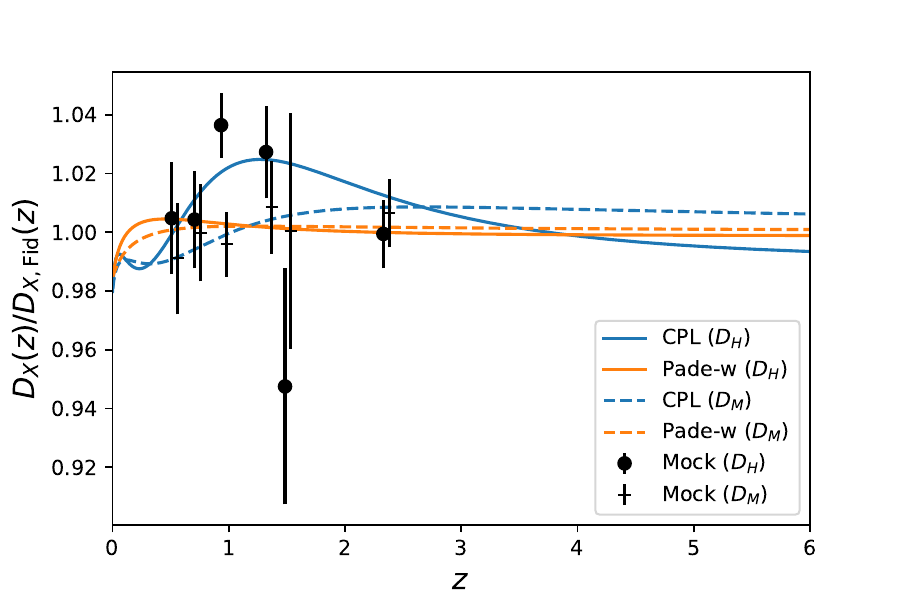}
    \caption{Best-fit curves for $D_H(z)$ (solid) and $D_M(z)$ (dashed) for both the CPL (blue) and Pade-$w$ (orange) models, all relative to those functions from the fiducial Pade-$w$ model.  The mock BAO data points are shown in black circles for the $D_H$ data points and pluses for the $D_M$ data points.  This specific mock dataset corresponds to the case where CPL is most preferred, with the greatest $\Delta \chi^2$.} 
    \label{fig:whycplisbetter}
\end{figure}

In Figure~\ref{fig:whycplisbetter} we show, proportional to the fiducial model, the functions $D_H(z)$ and $D_M(z)$ from both the best-fit CPL and Pade-$w$ models, fit to this same mock dataset that gives the largest spurious (since it is generated from the fiducial Pade-$w$ model) preference for CPL.  This kind of figure is useful for assessing from where this preference could arise.  We see that the black $D_H$ data points scatter high around $z\sim1$ while the $D_H(z)$ functions at lower redshift are closer to the fiducial values.  This kind of flexibility can only be achieved with DE being phantom above $z>0.5$, as allowed by CPL.

Furthermore, the fact that the high-z limit of the best-fit CPL and Pade-$w$ models tend to different values indicates that the two models prefer different values of $\Omega_{\rm m}h^2$. 
One way to prove Pade-$w$ (or any other purely quintessent, thawing DE model) is preferred over CPL is to better measure $\Omega_{\rm m}h^2$.  The current constraint on $\Omega_{\rm m}h^2$ comes from the CMB, which is cosmic variance limited. Thus, BAO or other $H(z)$ constraints at $z>3$ which are particularly useful for constraining $\Omega_{\rm m}h^2$ to higher precision, would allow differentiation of quintessent DE models from DE models with phantom crossing(s).

\section{Conclusions}

The combination of DESI Year 2 BAO measurements with Planck CMB data and multiple SN compilations provides growing evidence for evolving dark energy, with a preference for models in which the equation of state crosses the phantom divide line ($w = -1$). This crossing behavior, while interesting, raises theoretical challenges, as it is difficult to realize within standard physical frameworks such as scalar field models.

To assess the robustness of this preference, we conducted 1,000 Monte Carlo simulations using a phenomenological algebraic quintessence model, which provides a good fit to the real data, though not as favorable as the CPL form. Our analysis reveals that in over $30\%$ of the simulations, the CPL best-fit model exhibiting phantom crossing behavior performs better than the best-fit Pade-$w$ model. Most importantly, in $3.2\%$ of the simulations, the CPL model yields a $\Delta \chi^2$ (with respect to the Pade-$w$ model) larger than that observed in the real data.

In other words, the preference for phantom crossing in the real data can be at the $>2\sigma$ level, but not at the $5\sigma$ level required to be statistically conclusive.

This apparent preference may stem in part from the greater flexibility of the CPL parametrization and non-parametric methods, the uneven data distribution across redshift, and statistical fluctuations. This is not a criticism of the reliability of non-parametric methods nor the CPL parametrization in this particular case, but rather a reminder that $\Delta \chi^2$ values should be interpreted with care, when doing model comparison, to avoid misrepresenting the results. More precise and higher redsfhit BAO measurements will be useful in testing whether DE is purely quintessent or has a phantom crossing.

We should note that quintessence models that provide a good fit to the data, such as the best-fit Pade-$w$ model in the case of our analysis, require rapid variations in $w(z)$ and involve fine-tuning of parameters. Thus, both phenomenological models studied here present theoretical challenges that need to be addressed.

Importantly, we have not accounted for possible systematics in the data, so our results reflect only statistical variations. While the evidence for evolving dark energy is compelling, the specific nature of its behavior, especially the need for phantom crossing, remains uncertain. Future observations with improved precision and coverage will be critical to clarify these trends and test the consistency of various dark energy models.

\begin{acknowledgments}
We wish to acknowledge Eric Linder for comments in finalizing this draft.
\end{acknowledgments}

\bibliography{refs}

\end{document}